\documentclass[twoside,twocolumn,9pt]{article}
\usepackage{extsizes}
\usepackage[super,sort&compress,comma]{natbib} 
\usepackage[version=3]{mhchem}
\usepackage[left=1.5cm, right=1.5cm, top=1.785cm, bottom=2.0cm]{geometry}
\usepackage{balance}
\usepackage{mathptmx}
\usepackage{sectsty}
\usepackage{graphicx} 
\usepackage{lastpage}
\usepackage[format=plain,justification=justified,singlelinecheck=false,font={stretch=1.125,small,sf},labelfont=bf,labelsep=space]{caption}
\usepackage{float}
\usepackage{fancyhdr}
\usepackage{fnpos}
\usepackage[english]{babel}
\addto{\captionsenglish}{%
  
}
\usepackage{array}
\usepackage{droidsans}
\usepackage{charter}
\usepackage[T1]{fontenc}
\usepackage[usenames,dvipsnames]{xcolor}
\usepackage{setspace}
\usepackage[compact]{titlesec}
\usepackage[colorlinks=true,allcolors=blue]{hyperref}

\usepackage{overpic}
\usepackage{bm}
\usepackage[normalem]{ulem}
\usepackage{amssymb}
\usepackage{lmodern}
\usepackage[squaren]{SIunits}

\definecolor{cream}{RGB}{222,217,201}

\begin{document}

\pagestyle{fancy}
\thispagestyle{plain}
\fancypagestyle{plain}{
\renewcommand{\headrulewidth}{0pt}
}

\makeFNbottom
\makeatletter
\renewcommand\LARGE{\@setfontsize\LARGE{15pt}{17}}
\renewcommand\Large{\@setfontsize\Large{12pt}{14}}
\renewcommand\large{\@setfontsize\large{10pt}{12}}
\renewcommand\footnotesize{\@setfontsize\footnotesize{7pt}{10}}
\makeatother

\renewcommand{\thefootnote}{\fnsymbol{footnote}}
\renewcommand\footnoterule{\vspace*{1pt}%
\color{cream}\hrule width 3.5in height 0.4pt \color{black}\vspace*{5pt}} 
\setcounter{secnumdepth}{5}

\makeatletter 
\renewcommand\@biblabel[1]{#1}            
\renewcommand\@makefntext[1]%
{\noindent\makebox[0pt][r]{\@thefnmark\,}#1}
\makeatother 
\renewcommand{\figurename}{\small{Fig.}~}
\sectionfont{\sffamily\Large}
\subsectionfont{\normalsize}
\subsubsectionfont{\bf}
\setstretch{1.125} 
\setlength{\skip\footins}{0.8cm}
\setlength{\footnotesep}{0.25cm}
\setlength{\jot}{10pt}
\titlespacing*{\section}{0pt}{4pt}{4pt}
\titlespacing*{\subsection}{0pt}{15pt}{1pt}

\fancyfoot{}
\fancyfoot[LO,RE]{\vspace{-7.1pt}\includegraphics[height=9pt]{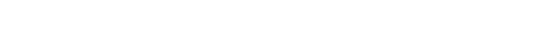}}
\fancyfoot[CO]{\vspace{-7.1pt}\hspace{13.2cm}\includegraphics{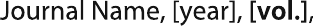}}
\fancyfoot[CE]{\vspace{-7.2pt}\hspace{-14.2cm}\includegraphics{head_foot/RF}}
\fancyfoot[RO]{\footnotesize{\sffamily{1--\pageref{LastPage} ~\textbar  \hspace{2pt}\thepage}}}
\fancyfoot[LE]{\footnotesize{\sffamily{\thepage~\textbar\hspace{3.45cm} 1--\pageref{LastPage}}}}
\fancyhead{}
\renewcommand{\headrulewidth}{0pt} 
\renewcommand{\footrulewidth}{0pt}
\setlength{\arrayrulewidth}{1pt}
\setlength{\columnsep}{6.5mm}
\setlength\bibsep{1pt}

\makeatletter 
\newlength{\figrulesep} 
\setlength{\figrulesep}{0.5\textfloatsep} 

\newcommand{\topfigrule}{\vspace*{-1pt}%
\noindent{\color{cream}\rule[-\figrulesep]{\columnwidth}{1.5pt}} }

\newcommand{\botfigrule}{\vspace*{-2pt}%
\noindent{\color{cream}\rule[\figrulesep]{\columnwidth}{1.5pt}} }

\newcommand{\dblfigrule}{\vspace*{-1pt}%
\noindent{\color{cream}\rule[-\figrulesep]{\textwidth}{1.5pt}} }

\definecolor{mmcolor}{rgb}{0.0, 0.0, 0.8}
\definecolor{deletecolor}{rgb}{0.8, 0.0, 0.0}
\newcommand{\mm}[2]{{\color{mmcolor}#1}{\color{deletecolor}\sout{#2}}}
\newcommand{\mme}[2]{{\color{mmcolor}#1}{\color{deletecolor}#2}}

\newcommand{\sg}[2]{{\color{red}#1}{\color{deletecolor}\sout{#2}}}

\makeatother

\twocolumn[
  \begin{@twocolumnfalse}
{\includegraphics[height=30pt]{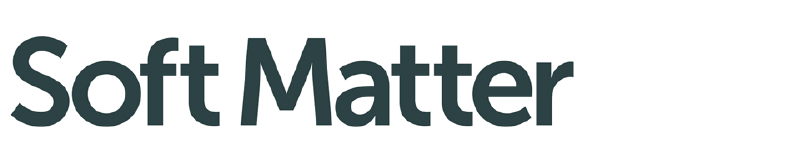}\hfill\raisebox{0pt}[0pt][0pt]{\includegraphics[height=55pt]{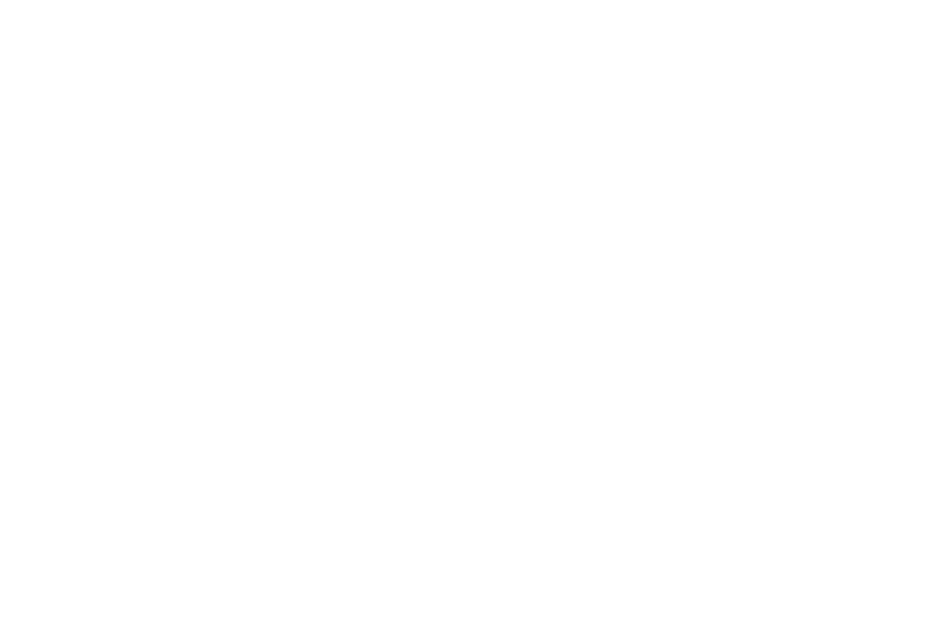}}\\[1ex]
\includegraphics[width=18.5cm]{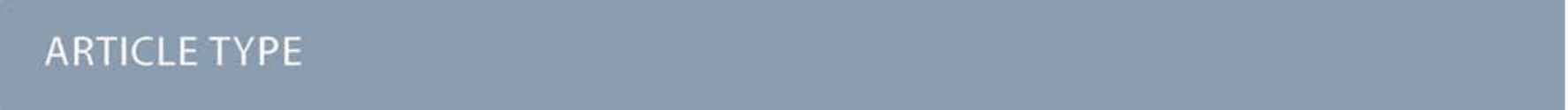}}\par
\vspace{1em}
\sffamily
\begin{tabular}{m{4.5cm} p{13.5cm} }

\includegraphics{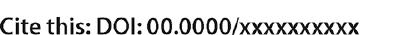} & \noindent\LARGE{\textbf{Phase separation dynamics in deformable droplets}} \\
\vspace{0.3cm} & \vspace{0.3cm} \\

 & \noindent\large{Simon Gsell,$^{\ast}$\textit{$^{a,b}$} and Matthias Merkel$^{\dagger}$\textit{$^{a}$}} \\

\includegraphics{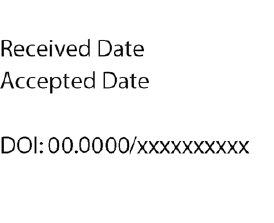} & \noindent\normalsize{%
  Phase separation can drive spatial organization of multicomponent mixtures.
  For instance in developing animal embryos, effective phase separation descriptions have been used to account for the spatial organization of different tissue types.
  Similarly, separation of different tissue types and the emergence of a polar organization is also observed in cell aggregates mimicking early embryonic axis formation.
  Here, we describe such aggregates as deformable two-phase fluid droplets, which are suspended in a fluid environment (third phase). 
  Using hybrid finite-volume Lattice-Boltzmann simulations, we numerically explore the out-of-equilibrium routes that can lead to the polar equilibrium state of such a droplet (Janus droplet).
  We focus on the interplay between spinodal decomposition and advection with hydrodynamic flows driven by interface tensions, which we characterize by a Peclet number $Pe$.
  Consistent with previous work, for large $Pe$ the coarsening process is generally accelerated.
  However, for intermediate $Pe$ we observe long-lived, strongly elongated droplets, where both phases form an alternating stripe pattern. We show that these ``croissant'' states are close to mechanical equilibrium and coarsen only slowly through diffusive fluxes in an Ostwald-ripening-like process.
  Finally, we show that a surface tension asymmetry between both droplet phases leads to transient, rotationally symmetric states whose resolution leads to flows reminiscent of Marangoni flows.
  Our work highlights the importance of advection for the phase separation process in finite, deformable systems.

} \\

\end{tabular}

 \end{@twocolumnfalse} \vspace{0.6cm}

  ]

\renewcommand*\rmdefault{bch}\normalfont\upshape
\rmfamily
\section*{}
\vspace{-1cm}


\footnotetext{\textit{$^{a}$~Aix Marseille Univ, Universit\'e de Toulon, CNRS, CPT (UMR 7332), Turing Centre for Living systems, Marseille, France}}
\footnotetext{\textit{$^{b}$~Aix Marseille Univ, CNRS, IBDM (UMR 7288), Turing Centre for Living systems, Marseille, France}}
\footnotetext{$^{\ast}$~{simon.gsell@univ-amu.fr}}
\footnotetext{$^{\dagger}$~{matthias.merkel@cnrs.fr}}






\section{Introduction}
\label{sec:intro}
Systems composed of two or more components can undergo phase separation.\cite{bray02,mao19,mao20}
During this process, phase domains of different composition first emerge either through spinodal decomposition or nucleation, before these domains coarsen through diffusive processes such as Ostwald ripening.  
Coarsening is associated with a power-law growth of typical domain size $\lambda$, which scales as $\lambda\sim t^{1/3}$ with time $t$.\cite{bray02}
In the case of fluid-fluid phase separation, coarsening can also occur through advection with hydrodynamic flows that are driven by the interface tension between adjacent domains.
Such hydrodynamic flows can speed up domain coarsening, which can lead to e.g.\ a linear scaling $\lambda\sim t$.\cite{bray02}
However, the precise interplay between diffusive and hydrodynamic processes can be quite complex, leading to a rich phenomenology.\cite{Tanaka1998}

While phase separation has long been employed for industrial processes, it has also been identified as a possible key organizing mechanism in biological systems, from the sub-cellular\cite{hyman14} to the ecological\cite{liu13} scale.
For example, the interior of biological cells is organized in part through fluid compartments that can emerge from phase separation.\cite{brangwynne13,hyman14,wheeler18}
Phase separation has moreover been used as effective model to describe the formation of multi-cellular aggregates.\cite{Kuan2021}
But it has also been used to describe separation of different types of biological tissues.\cite{steinberg63,foty05,cachat16,Amack2012}
This was motivated by experiments showing that in many animal species early differentiated cell populations unmix.\cite{Townes1955,Schoetz2008}

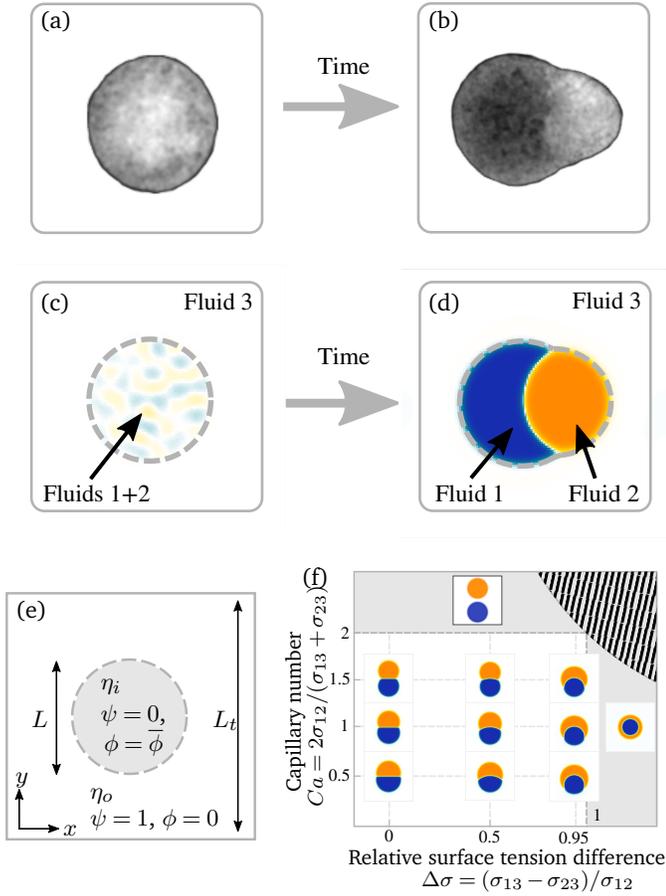
\begin{figure}[!t]
  \begin{center}
    \begin{overpic}[width=0.47\textwidth]{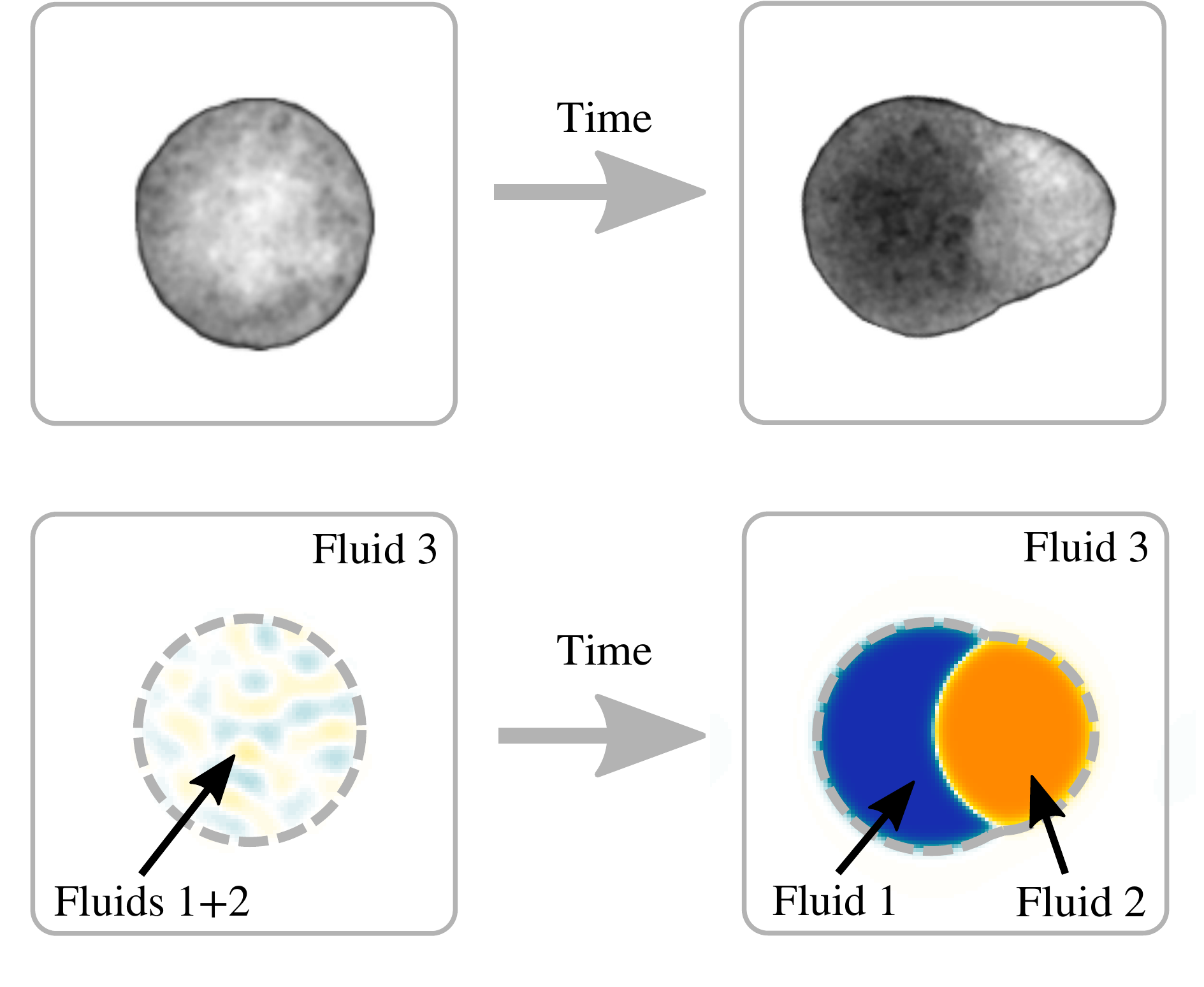}
      \put(4,78){(a)}
      \put(63,78){(b)}
      \put(4, 35){(c)}
      \put(63,35){(d)}
    \end{overpic}
    \vskip1em
    \begin{overpic}[width=0.18\textwidth,trim=0cm -0.5cm 0cm 0cm, clip=True]{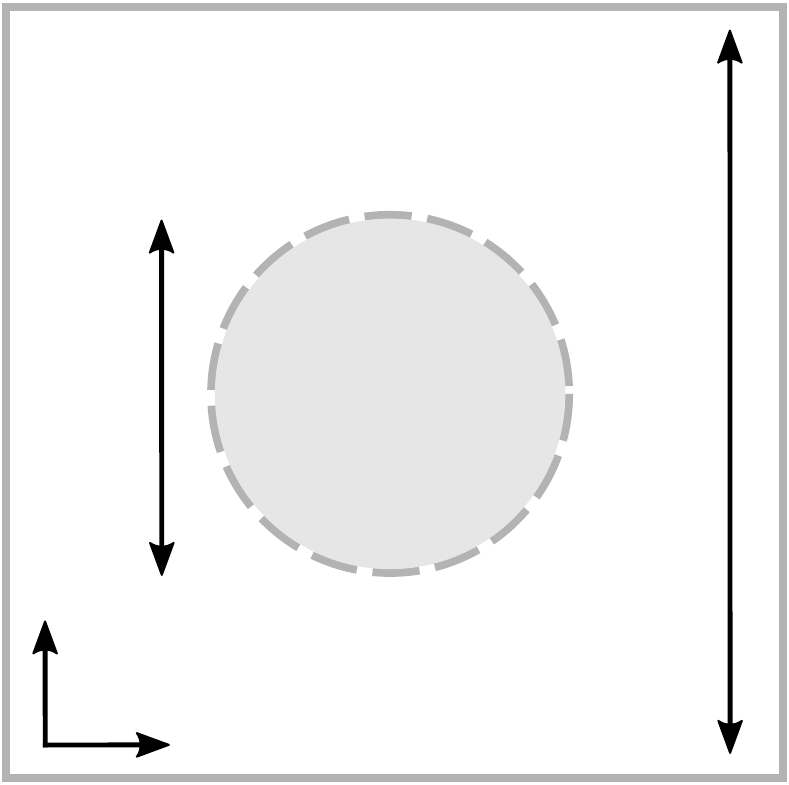}
      
      \put(4,90){(e)}
      
      \put(10,49){$L$}
      \put(78,49){$L_t$}
      \put(22,8){$x$}
      \put(4,28){$y$}
      
      \put(36,63){$\eta_i$}
      \put(32,22){$\eta_o$}
      \put(36,40){\shortstack{$\psi=0$, \\[-0.2em] $\phi = \overline{\phi}$}}
      \put(32,12){$\psi=1$, $\phi = 0$}
      
    \end{overpic}
    \hskip2.3em
    \begin{overpic}[width=0.26\textwidth]{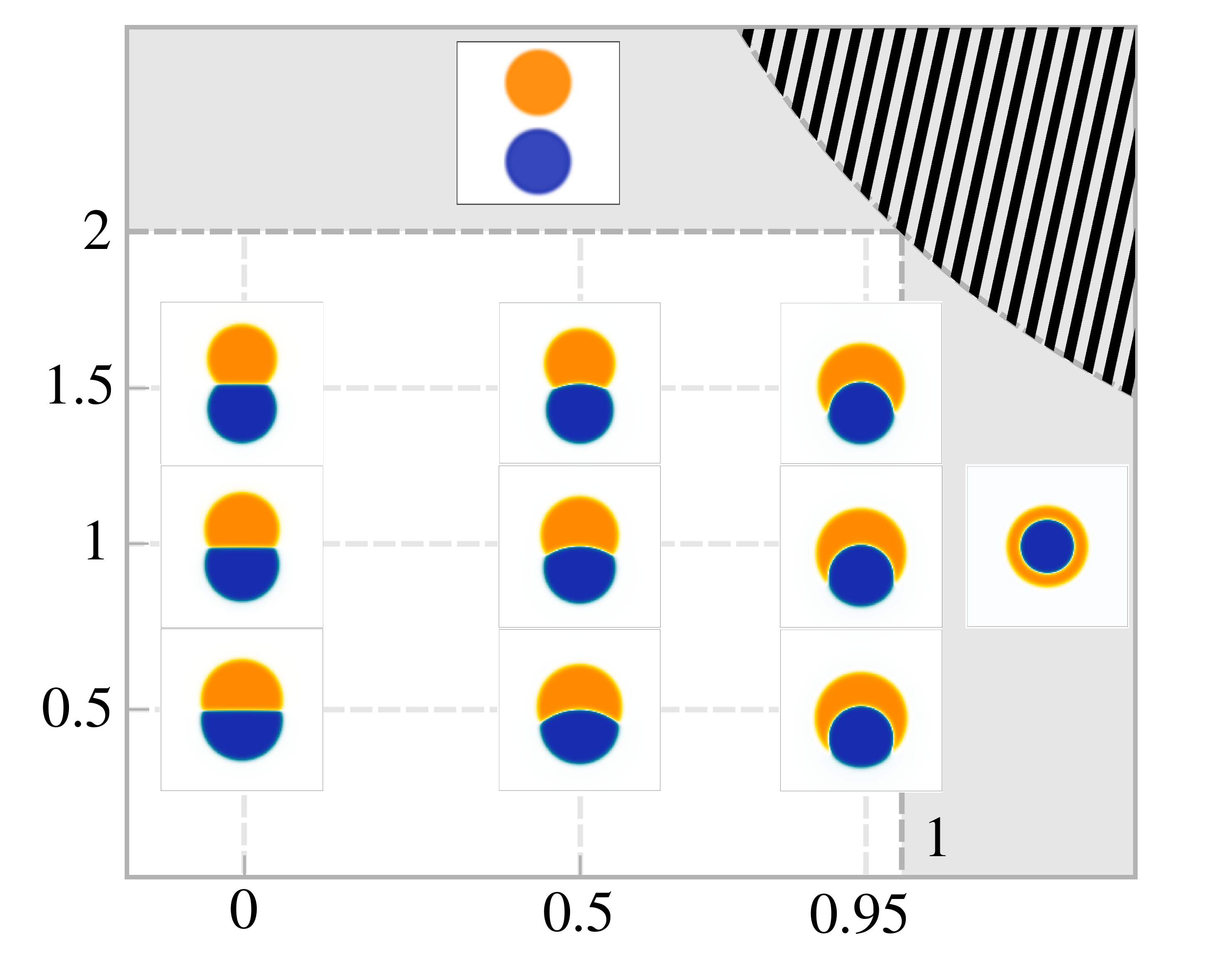}
      \put(-4,75){(f)}
      \put(9,-2){\small Relative surface tension difference}
      \put(29,-8){\small $\Delta \sigma = (\sigma_{13} - \sigma_{23}) / \sigma_{12}$}
      \put(-8.5,18){\rotatebox{90}{\small Capillary number}}
      \put(-2.5,14){\rotatebox{90}{\small $Ca = 2 \sigma_{12} / (\sigma_{13} + \sigma_{23})$}}
    \end{overpic}
  \end{center}
  \caption{Phase separation and symmetry breaking in deformable multicomponent droplets. 
  (a,b) Microscopy images of cell aggregates mimicking early vertebrate development, taken 1 day apart from each other. Brighter regions indicate the early mesoderm marker T/Brachyury (images taken from Ref.~\cite{Hashmi2020}).
  (c-e) Ternary fluid mixture model: a droplet, immersed in fluid 3 and initially composed of a mixture of fluids 1 and 2 (c), undergoes phase-separation until reaching the thermodynamic equilibrium state (d) characterized by a surface tension balance between all interfaces.
  (e) Schematic representation of the physical configuration.
  (f) Dependence of the thermodynamic equilibrium state on a Capillary number $Ca = 2 \sigma_{12} / (\sigma_{13} + \sigma_{23})$ and the relative surface tension difference $\Delta \sigma = (\sigma_{13} - \sigma_{23}) / \sigma_{12}$.
  The hatched region is unphysical, because at least one of the tensions $\sigma_{ij}$ would need to be negative there.
  }
  \label{fig:intro}
\end{figure}

Phase separation of biological tissues may also help to understand an important question in the field of developmental biology: How does the vertebrate axis form during animal development?
To address this and related questions, experimentalists study aggregates of stem cells, called gastruloids, which have been shown to mimic vertebrate development (Fig.~\ref{fig:intro}a,b).\cite{VandenBrink2014,Beccari2018,Fulton2020}
Such gastruloids initially consist of spherical aggregates of undifferentiated stem cells (Fig.~\ref{fig:intro}a).  Even under isotropic boundary conditions, these aggregates then reliably\cite{Cermola2021} form polarized structures with different cell populations (Fig.~\ref{fig:intro}b).\cite{Hashmi2020}
This axis later defines the axis of the vertebral column.\cite{Beccari2018,VandenBrink2020}
However, how exactly this axis initially emerges has remained unclear.

As a first step towards building a better physical understanding of such cellular aggregates, we study here the passive dynamics of a ternary phase separating system (Fig.~\ref{fig:intro}c-e), where a droplet made out of two fluid phases (1 and 2) is suspended in a third fluid phase (3).
Note that such multi-component emulsion droplets\cite{utada05,nisisako16,wang20} are also important in many other contexts, including the food and pharmaceutical industries.\cite{iqbal15}
There is a limited number of possibilities for the thermodynamic equilibrium state of such a ternary system, depending on the ratios of the interface tension $\sigma_{12}$ to the surface tensions $\sigma_{13}$ and $\sigma_{23}$ (Fig.~\ref{fig:intro}f).  
Here, we focus on the case where the equilibrium state is a Janus droplet (Fig.~\ref{fig:intro}d), i.e.\ where $\sigma_{12}<\sigma_{13}+\sigma_{23}$ and $|\sigma_{13}-\sigma_{23}|<\sigma_{12}$ (white region in Fig.~\ref{fig:intro}f). 
While the equilibrium state is well known, we are interested in the dynamics that leads to this state, and how it depends on the interplay between diffusion and hydrodynamic flows. 
We start from an initial state where fluids 1 and 2 are almost homogeneously mixed inside of a spherical droplet (Fig.~\ref{fig:intro}c). 
Thus, during the phase separation process, the system needs to spontaneously break rotational symmetry to reach the polarized equilibrium state (Fig.~\ref{fig:intro}d).

Here, we numerically study the dynamics of this ternary system using 2D hybrid finite-volume Lattice-Boltzmann simulations of coupled Cahn-Hilliard and Stokes equations.
We study the system's dynamics in terms of a Peclet number $Pe$ that characterizes the magnitude of hydrodynamic advection as compared to diffusive effects.
We find that while increasing $Pe$ generally speeds up the phase separation process, intermediate $Pe$ give rise to long-lived ``croissant'' states that slow down the relaxation to the final polarized state.
These results are not fundamentally changed by including an asymmetry between the surface tensions $\sigma_{13}$ and $\sigma_{23}$.
Taken together, our work demonstrates the importance of hydrodynamic flows in deformable multi-phase droplets.


\section{Model}
\label{sec:method}


\subsection{Free energy}
\label{sec:thermodynamics}
We describe a 2D ternary system where the three components have area fractions $C_1$, $C_2$, and $C_3$, which sum up to one everywhere:
\begin{equation}
 C_1 + C_2 + C_3 = 1. \label{eqn:sum_volume_fractions}
\end{equation}
The phase-separating behavior of these three components is described by the following free energy:\cite{semprebon16}
\begin{equation}
  \begin{split}
  \mathcal{F} = \int_\Omega \bigg[ &\dfrac{\kappa_1}{2} C_1^2 (1-C_1)^2 + \dfrac{\kappa_2}{2} C_2^2 (1-C_2)^2 + \dfrac{\kappa_3}{2} C_3^2 (1-C_3)^2 \\
  &+ \dfrac{\kappa_1^\prime}{2} (\nabla C_1)^2  + \dfrac{\kappa_2^\prime}{2} (\nabla C_2)^2  + \dfrac{\kappa_3^\prime}{2} (\nabla C_3)^2 \bigg] \mathrm{d}A.
  \end{split}
  \label{eqn:free_energy}
\end{equation}
The integral is over the total system area $\Omega$. For each component $j\in\lbrace 1,2,3\rbrace$, the free energy thus includes a quartic double-well potential with local minima at $C_j=0$ and $C_j=1$, and an energy density scale $\kappa_j$. Moreover, gradient terms with coefficients $\kappa'_j$ ensure finite widths of all interfaces.  
For any binary interface between phases of dominating components $j$ and $k$, i.e.\ where the third area fraction vanishes, interface width $\alpha_{jk}$ and interface tension $\sigma_{jk}$ take on the well-known values:\cite{semprebon16}
\begin{equation}
  \alpha_{jk} = \sqrt{\frac{ \kappa_j^\prime + \kappa_k^\prime }{ \kappa_j + \kappa_k } } \qquad\text{and}\qquad
  \sigma_{jk} = \dfrac{\alpha_{jk}}{6} (\kappa_j + \kappa_k).
\end{equation}
In the following, we will use the same width $\alpha=\alpha_{jk}$ for all interfaces $jk\in\lbrace 12, 13, 23\rbrace$.

\subsection{Dynamic equations}
\label{sec:dynamics}
To define the dynamics of the area fractions $C_j$ such that Eq.~\eqref{eqn:sum_volume_fractions} remains fulfilled, we follow Semprebon et al.\cite{semprebon16} by introducing phase fields $\phi$ and $\psi$ as
\begin{equation}
  \phi = C_1 - C_2 \quad\text{and}\quad  \psi = C_3.
\end{equation}
Using these definitions together with Eq.~\eqref{eqn:sum_volume_fractions}, the free energy in Eq.~\eqref{eqn:free_energy} can be expressed in terms of $\phi$ and $\psi$ instead of the $C_j$.
We then choose Cahn-Hilliard dynamics for both $\phi$ and $\psi$:
\begin{subequations}
\begin{align}
  \dfrac{\partial \phi}{\partial t} + \nabla \cdot ( \bm{u} \phi) &= \nabla \cdot \Big( [1-\psi] \Gamma_\phi \nabla \mu_\phi \Big)
  \label{eqn:CH_phi} \\
  \dfrac{\partial \psi}{\partial t} + \nabla \cdot ( \bm{u} \psi) &= \nabla \cdot \left( \Gamma_\psi \nabla \mu_\psi\right),
  \label{eqn:CH_psi}
\end{align}
\end{subequations}
where $t$ is time, $\bm{u}$ is the local fluid velocity, $\Gamma_\phi$ and $\Gamma_\psi$ are mobility parameters, and $\mu_\phi = \delta \mathcal{F} / \delta \phi$ and $\mu_\psi = \delta \mathcal{F} / \delta \psi$ are the chemical potentials of $\phi$ and $\psi$, respectively.
The factor $[1-\psi]$ in Eq.~\eqref{eqn:CH_phi} ensures that fluids 1 and 2 do not diffuse across fluid 3.  This is motivated by the biological background -- we exclude any dissociation and re-association of cells (components 1 and 2) from/to the aggregate.

To account for hydrodynamic effects, we describe all phases of the system as incompressible Newtonian fluids, using the incompressible Stokes' equations:
\begin{subequations}
\begin{align}
  0 &= \nabla \cdot \bm{u}, \label{eqn:divU}\\
  0 &= - \nabla \Pi + \nabla \cdot \left( 2 \eta \bm{S} \right) - \phi \nabla \mu_\phi - \psi \nabla \mu_\psi. \label{eqn:NS}
\end{align}
\end{subequations}
Here, $\Pi$ is the fluid pressure, $\eta$ is the fluid viscosity, $\bm{S}$ is the shear-rate tensor $\bm{S} = (\nabla \bm{u} + (\nabla \bm{u})^T)/2$, and the last two terms correspond to the forces induced by the scalar fields.\cite{bray02}
We focus here on overdamped dynamics, as this is the relevant limit for $\sim 100\micro\meter$-scale cellular aggregates with shear rates on the $\sim\text{day}^{-1}$ scale.
Because the effective viscosity of embryonic cell aggregates is typically many orders of magnitude above that of water,\cite{David2014} we choose a $\psi$-dependent viscosity $\eta(\psi) = (1-\psi)\eta_i + \psi \eta_o$, where $\eta_i\gg\eta_o$.

\subsection{Initial and boundary conditions}
\label{sec:ICs_and_BCs}
We initialize the 2D system in a configuration schematized in figure~\ref{fig:intro}c,e.
The system initially consists of a circular droplet with diameter $L$, composed of a mixture of fluids 1 and 2 with $\psi = 0$, which is surrounded by fluid 3 with $\psi=1$ and $\phi=0$.
The inside of the droplet is initialized as $\phi=\bar{\phi}+\Delta\phi_{IC}\xi$, where $\bar\phi$ and $\Delta\phi_{IC}$ are scalar parameters, and $\xi$ is zero-mean, delta-correlated Gaussian white noise. 
The parameter $\bar\phi$ corresponds to the conserved area average of $\phi$, defined as $\bar\phi = \int_{\Omega} (1-\psi)\phi\,\mathrm{d}A/\int_{\Omega} (1-\psi)\,\mathrm{d}A$. 
We simulate the system in a squared periodic box with side length $L_t$.  Parameter values are listed in table~\ref{tab:parameter_values}.
\subsection{Dimensionless parameters}
\label{sec:parameters}
\subsubsection{Equilibrium state}
The equilibrium state depends on four dimensionless parameters.  These are first, the difference $\bar\phi$ in area fractions between fluids 1 and 2. Here we focus exclusively on the case $\bar\phi=0$, i.e.\ each fluid occupies half of the droplet area. Second, we set the ratio between droplet size $L$ and interface width $\alpha$ to $L/\alpha=64$.

Finally, the equilibrium state also depends on the three tensions $\sigma_{jk}$, for which we choose the following two dimensionless ratios:
\begin{equation}
  Ca = \dfrac{2\sigma_{12}}{\sigma_{13} + \sigma_{23}},\qquad \Delta \sigma = \dfrac{\sigma_{13} - \sigma_{23}}{\sigma_{12}}. 
\end{equation}
The parameter $Ca$ is akin to a Capillary number -- this ratio between the inner and outer surface tensions controls the deformability of the droplet due to capillary effects. Unless stated otherwise, we set $Ca=1$.
The parameter $\Delta \sigma$ quantifies the surface tension difference between fluids 1 and 2 with respect to fluid 3. 
In the following, $\Delta \sigma$ is varied over the range $0 \leq \Delta \sigma < 1$. 
We are thus in the regime where the equilibrium state is a dipole structure (Fig.~\ref{fig:intro}f).

\subsubsection{Relaxation dynamics}
The four phenomenological coefficients $\Gamma_\phi$, $\Gamma_\psi$, $\eta_i$, and $\eta_o$ affect only the relaxation dynamics. Absorbing one to rescale time, we obtain three dimensionless parameters.

First, we define a Peclet number that compares advective fluxes due to hydrodynamic flows to the diffusive fluxes of the $\phi$-field:
\begin{equation}
  Pe = \frac{L \alpha}{\Gamma_\phi \eta_{i}}.
\end{equation}
This number is defined on the scale of the droplet size $L$; it corresponds to the ratio $Pe=\tau_D/\tau_A$ between a diffusive time scale $\tau_D = L^2/D_\phi$ with the typical diffusion coefficient scale $D_\phi=\Gamma_\phi \sigma_{12}/\alpha\sim\Gamma_\phi(\kappa_1+\kappa_2)$, and an advective time scale $\tau_A = L/v_f$ with the typical flow velocity scale $v_f=\sigma_{12}/\eta_i$. Increasing the Peclet number corresponds to increasing the influence of advection.  
The limit case $Pe=0$ is simulated by ignoring the advection term in equations (\ref{eqn:CH_phi})-(\ref{eqn:CH_psi}). In this case, we set $\Gamma_\phi/\Gamma_\psi = 3$.
In this article, we will vary $Pe$ to probe how the interplay of diffusion and hydrodynamic flows affects the phase separation dynamics.

Second, we define a Peclet number with respect to the $\psi$ field, and with respect to the interface width $\alpha$, as $Pe_\psi=\alpha^2/(\Gamma_\psi \eta_{i})$.
Here, we fix this number to a relatively small value of $Pe_\psi=15/32$ to ensure that the outer droplet interface remains stable during the simulations.

Third, we fix the viscosity ratio between droplet and the surrounding fluid to the value $\eta_i/\eta_o=20$. 
We have checked that changes of this ratio do not strongly affect the simulation results as long as $\eta_i/\eta_o\gtrsim 10$.

\begin{table}
  \small
  \caption{Values of dimensionless parameters used in our simulations
  \label{tab:parameter_values}}
  \begin{tabular}{lccccccc}
    \hline \\[-8pt]
    parameter  & $\bar\phi$ & $L/\alpha$ & $Ca$ & $Pe_\psi$ & $\eta_i/\eta_o$ & $\Delta\phi_{IC}$ & $L_t/L$ 
    \\\hline\\[-8pt]
    value     & $0$ & $64$ & $1$ & $15/32$ & $20$ & $0.01$ & $2$ 
    \\\hline
  \end{tabular}\\[8pt]
  \begin{tabular}{lcc}
    \hline \\[-8pt]
    parameter  & $\Delta x/\alpha$ & $\Delta t/(\Delta x\,\eta_i/\sigma_{12})$
    \\\hline\\[-8pt]
    value     & $1$ & $1/640$
    \\\hline
  \end{tabular}
\end{table}

\begin{figure*}[!t]
  \includegraphics[width=\textwidth]{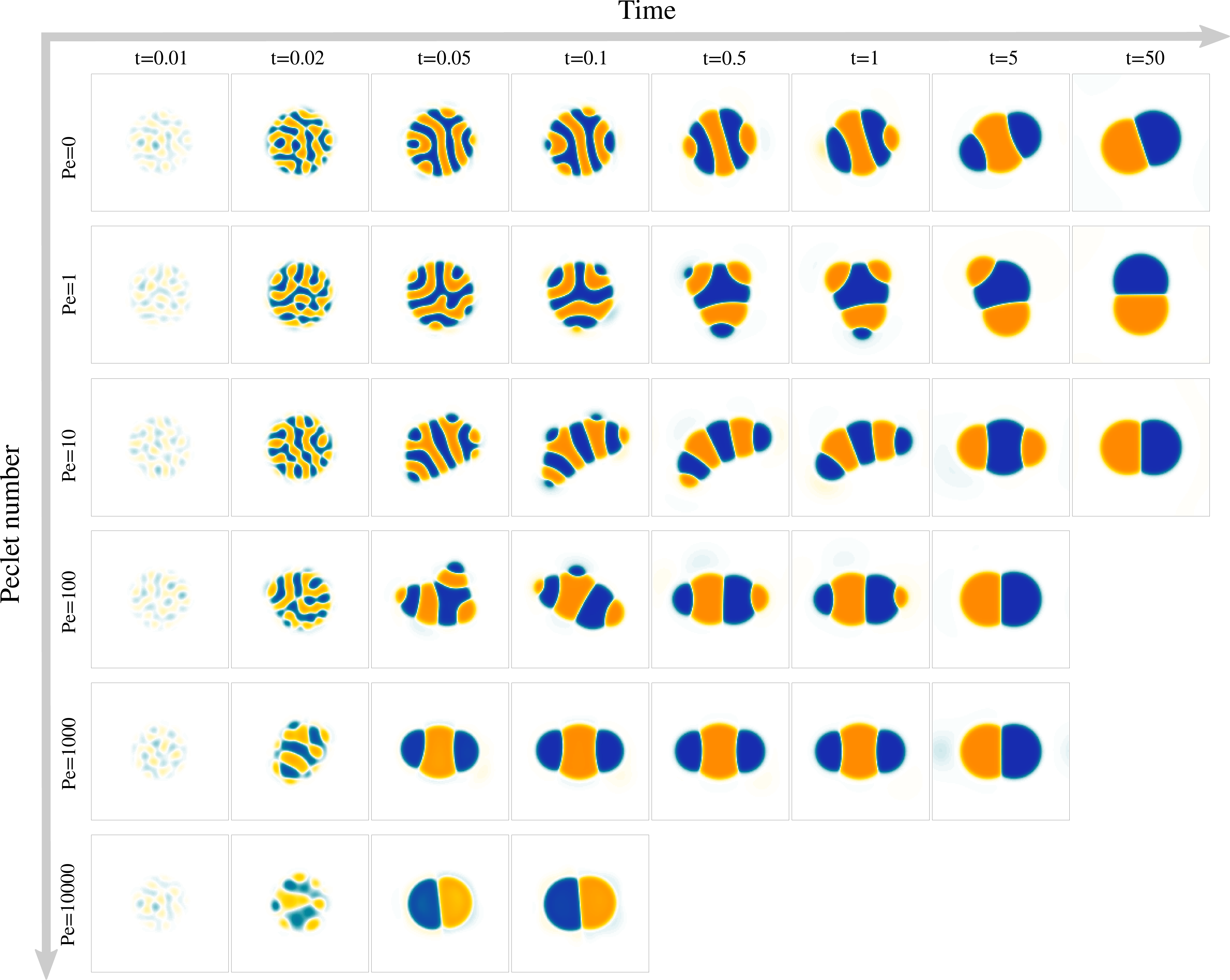}
  \caption{Advection affects the coarsening dynamics and the time to reach the polarized equilibrium state. 
  For each value of the Peclet number, a time series of snapshots is shown, representing the $\phi$ field from $-1$ (blue) to $1$ (orange).
  Here we set $\Delta \sigma = 0$, i.e.\ the outer surface tension is independent of $\phi$.
  }
  \label{fig:snaps_Pe}
\end{figure*}

\subsection{Numerical implementation}
\label{sec:num}
We solve the dynamical equations \eqref{eqn:CH_phi}--\eqref{eqn:NS} using a hybrid finite-volume Lattice Boltzmann approach.
Similar methods have been used in earlier publications,\cite{tiribocchi09, blow14, doostmohammadi16}
using finite-difference schemes for the phase field dynamics. Here, we use a finite-volume scheme instead to ensure exact conservation of the scalar fields $\phi$ and $\psi$.\cite{eymard00}

We discretize space using a Cartesian square grid with spacing $\Delta x=\alpha$.  For the advection terms in Eqs.~\eqref{eqn:CH_phi} and \eqref{eqn:CH_psi}, we use a first-order upwind scheme, and for the diffusive fluxes we use a central second-order scheme. 
We integrate over time using a first-order explicit Euler scheme. We solve the incompressible Stokes equations \eqref{eqn:divU} and \eqref{eqn:NS} by including inertial terms with a small Reynolds number, which we integrate using a two-relaxation-time Lattice Boltzmann method.\cite{gsell21}
The time step for both finite-volume and Lattice Boltzmann parts is $\Delta t=(\Delta x/v_f)/640$ with the flow velocity scale $v_f=\sigma_{12}/\eta_i$.
Details on the numerical method are provided in the Supplementary Information.


\section{Results}
\label{sec:results}
In the following, we examine how the droplet polarization dynamics depends (i) on the magnitude of advection as quantified by the Peclet number $Pe$ (sections~\ref{sec:Pe} and \ref{sec:elong}) and (ii) on surface tension asymmetry, $\Delta\sigma\neq 0$ (section~\ref{sec:dsig}).

\subsection{Advection speeds up the polarization process}
\label{sec:Pe}
To study the role of advection for the phase separation process, we run simulations with symmetric surface tensions $\Delta\sigma=0$ and varying Peclet number $Pe$ (Fig.~\ref{fig:snaps_Pe}, Movies S1-S3). We find well-known features of spinodal decomposition, including an initial decomposition phase where $|\phi|$ generally grows until it saturates at $\approx 1$, and a subsequent coarsening phase.
However, Fig.~\ref{fig:snaps_Pe} also suggests (1) that increasing the Peclet number speeds up the phase separation process (discussed below in this section),\cite{bray02} and (2) that intermediate Peclet numbers give rise to long-lived strongly elongated droplets (discussed in the next section, \ref{sec:elong}).

\begin{figure}[!t]
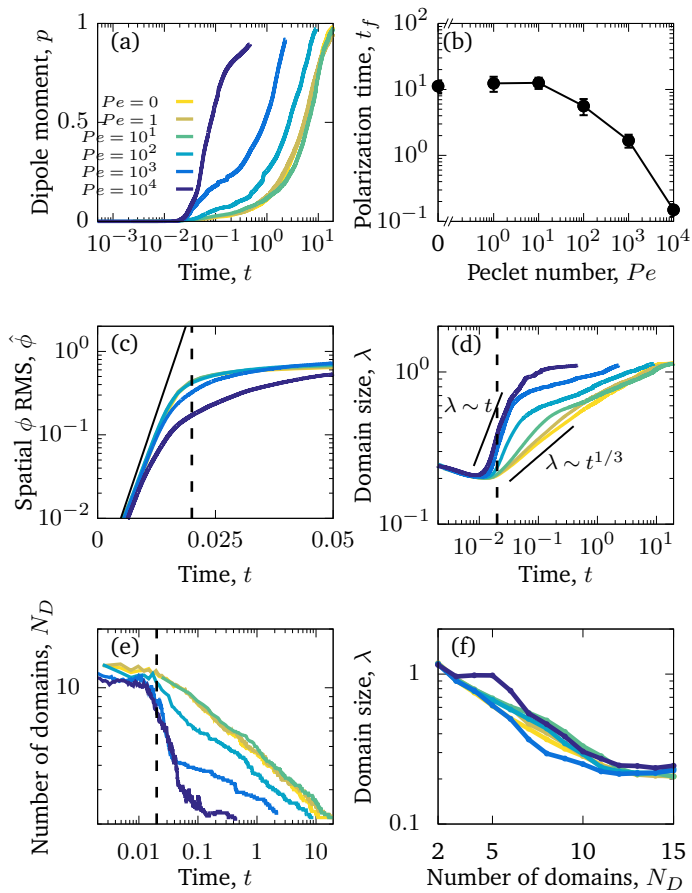

  \begin{center}

    \input{single/dipole_Pe.tex}
    \input{single/tf_Pe.tex}
    \input{single/decomp_Pe.tex}
    \input{single/lambda_Pe.tex}
    \input{single/domains_Pe.tex}
    \input{single/lambda_domains_bin.tex}

  \end{center}
  \caption{Advection speeds up droplet polarization by accelerating the coarsening phase.
  (a) Dipole moment $p$ depending on time for different $Pe$.
  (b) Polarization time $t_p$ as a function of the Peclet number; error bars indicate the standard error of the mean.
  (c) Spatial root-mean-square of $\phi$ depending on time. The black line shows the initial growth with rate $\omega$ predicted by linear stability analysis, $\hat{\phi} \sim e^{\omega t}$.
  (d) Domain size $\lambda$ as defined in the SI depending on time.
  (e) Number of phase domains $N_D$ depending on time.
  (f) Domain size $\lambda$ as a function of the number of phase domains $N_D$.
  Legends in panels c-f same as in panel a.
  In panels c-e, the vertical dashed line indicates the approximate end time $t_d \approx 0.02 \tau_D$ of the initial decomposition phase.
  The quantities on all ordinate axes are averaged over 25 simulation runs. 
  }
  \label{fig:speedup_Pe}
\end{figure}

To quantify in how far increasing the Peclet number speeds up the droplet polarization process, we measure the time evolution of a dipole moment, which we define as
\begin{equation}
  \bm{P}(t) = \int_\Omega (1 - \psi)\phi\,(\bm{x}-\bm{c}) \,\mathrm{d}A.
\end{equation}
Here, $\bm{c}$ is the barycenter of the droplet, defined as $\bm{c} = \int_\Omega (1 - \psi) \bm{x} \,\mathrm{d}A / \int_\Omega (1 - \psi) \,\mathrm{d}A$. 
We study the normalized magnitude $p=|\bm{P}|/P_0$, where $P_0 = L^3 / 6$ is a reference dipole moment, which corresponds to a circular droplet of diameter $L$ split in two semi-circular regions with $\phi = 1$ and $\phi = -1$.
We find that indeed, the dipole moment $p$ grows faster for higher Peclet number (Fig.~\ref{fig:speedup_Pe}a).
We moreover define a polarization time $t_p$ of the system as the time when the polarization first reaches $p\geq0.9p_\mathrm{eq}$, where $p_\mathrm{eq}$ is the dipole moment of the equilibrium state.  Consistent with known results from the literature,\cite{bray02} we find that advection speeds up the phase separation process (Fig.~\ref{fig:speedup_Pe}b). However, we note this effect only for $Pe \gtrsim 100$, while $t_p$ remains approximately constant for Peclet numbers below 100. 
To better understand this $Pe$ dependence of the polarization process, we now discuss the different phases of the process separately.

\textbf{Initial decomposition.}
We quantify the phase field amplitude $|\phi|$ during the initial decomposition phase, which lasts until the time $t_d\approx0.02\tau_D$.
To this end, we use the spatial root mean square $\hat\phi$ (RMS) of $\phi$, defined as $\hat\phi^2=\int_\Omega (1 - \psi)\phi^2 \,\mathrm{d}A/\int_\Omega (1 - \psi)\,\mathrm{d}A$.  Fig.~\ref{fig:speedup_Pe}c shows that $\hat\phi$ initially increases exponentially, and reaches a saturation at approximately $t_d$. 
Except for very large $Pe$, the behavior during this initial phase does not depend very strongly on the Peclet number.  

\textbf{Coarsening.}
We find that the subsequent coarsening phase is strongly affected by the Peclet number. This is apparent in Fig.~\ref{fig:speedup_Pe}d, where we plot the time evolution of the typical domain size $\lambda$ (exact definition in SI).
For $Pe=0$, the domain size $\lambda$ coarsens according to an approximately constant power law with an exponent close to $1/3$. In 2D, this is indeed the predicted exponent for systems that coarsen exclusively through the motion and deformation of domain boundaries,\cite{bray02} even though we also observe occasional domain evaporation (Fig.~\ref{fig:snaps_Pe}).

For very large Peclet number, $Pe\gtrsim 10^3$, we observe a coarsening exponent close to one (Fig.~\ref{fig:speedup_Pe}d), which corresponds to the expected result for advection-dominated coarsening. For such large $Pe$ numbers, advection even induces noticeable coarsening already during the initial decomposition phase (Fig.~\ref{fig:speedup_Pe}d). This early coarsening is also the likely reason for the slowing down of the initial decomposition for very large $Pe$ (Fig.~\ref{fig:speedup_Pe}c), since diffusive unmixing needs to act over larger length scales in this regime.

The difference in the coarsening behavior between small and large Peclet number is also apparent when studying the behavior of the number of phase domains $N_D$ (Fig.~\ref{fig:speedup_Pe}e).
We computationally define a phase domain as a connected region with constant sign of $\phi$. 
The number of domains $N_D$ decreases during the coarsening process until it reaches a minimum of two, which is attained in the equilibrium state.
Indeed, $N_D$ decreases more rapidly for larger Peclet number Fig.~\ref{fig:speedup_Pe}e. This confirms that advection speeds up the coarsening process, as $N_D$ and $\lambda$ are strongly correlated (Fig.~\ref{fig:speedup_Pe}f). 

For intermediate Peclet number, we initially observe a relatively fast coarsening (Fig.~\ref{fig:speedup_Pe}d). However, coarsening subsequently significantly slows down, which is also the reason why the polarization time $t_p$ decreases only for very large $Pe$ (compare Figs.~\ref{fig:speedup_Pe}b,d).
What could be responsible for this slow down?


\subsection{Elongated, striped droplets at intermediate $Pe$}
\label{sec:elong}

\begin{figure*}[t!]
  \begin{center}

    \begin{overpic}[width=0.3\textwidth, trim=0cm -1.cm 0cm 0cm]{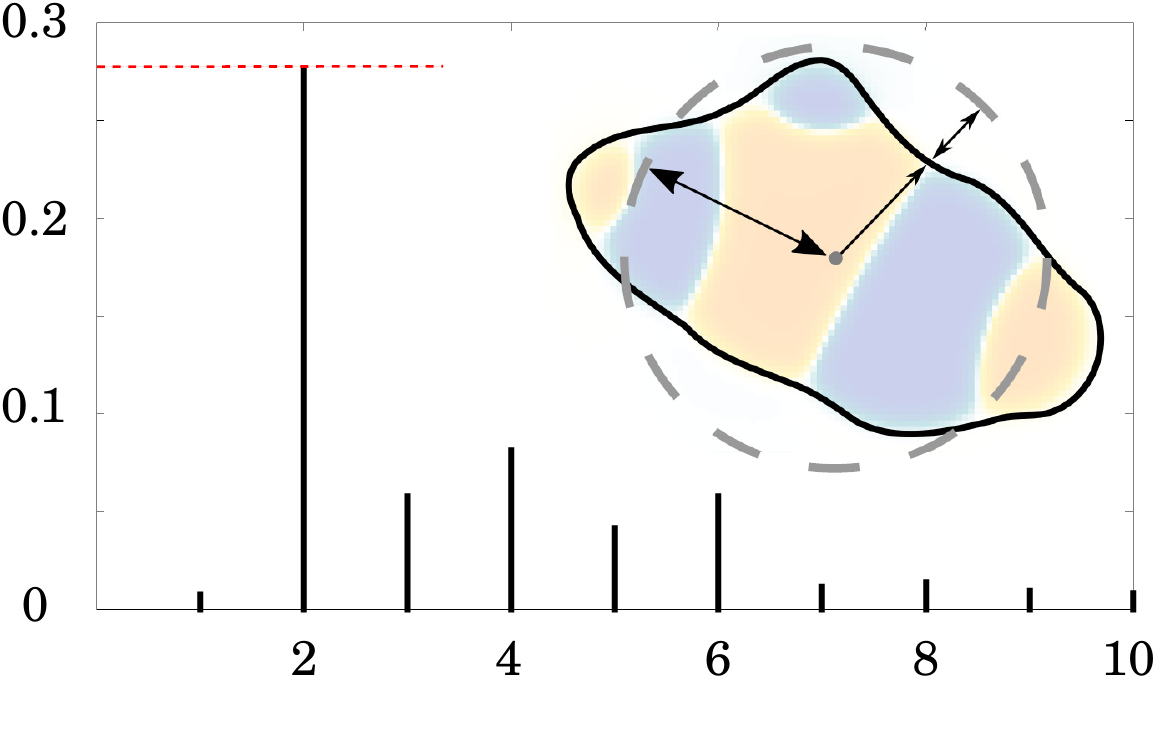}
      \put(0,80){(a)}
      \put(71,47){$\bm{c}$}
      \put(58,48){$L / 2$}
      \put(76,52){$r(\theta)$}
      \put(85,64){$\delta(\theta)$}
      \put(-7,23){\rotatebox{90}{Fourier amplitude}}
      \put(25,10){Angular wave number}
      \put(40,66){\color{red}$\bm{E}$}
    \end{overpic}
    ~~~~~
    \input{single/I2_inter_Pe.tex}
    \input{single/A2_Pe.tex}    
    
    \begin{overpic}[width=0.3\textwidth]{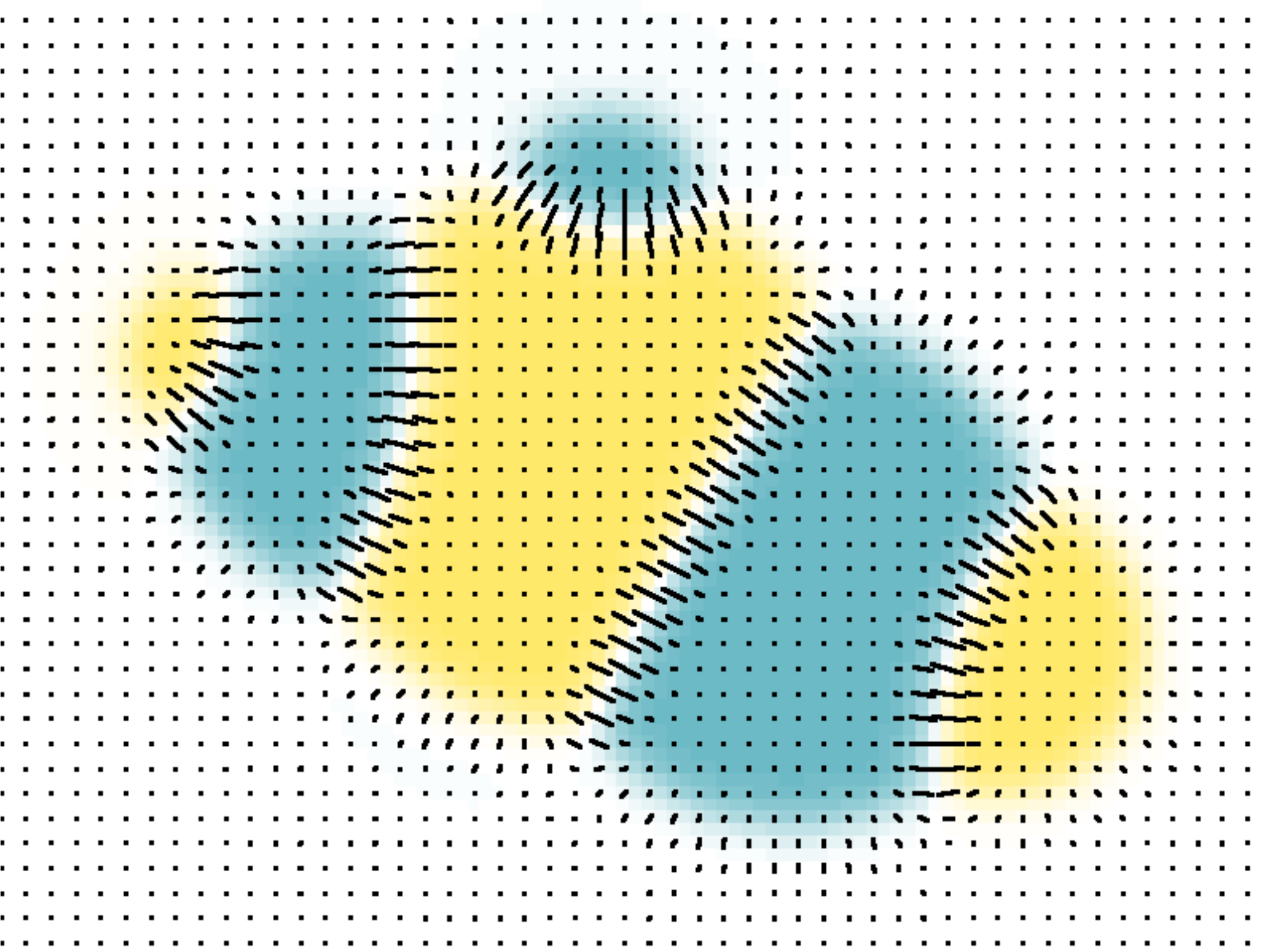}
      \put(0, 75){(d)}
      \put(55,75){$Pe=100$, $t=0.1$ }
    \end{overpic}
    \input{single/nematic_Pe.tex}
    \input{single/nematic_domains_Pe_bin.tex}

    \vskip1em
    \begin{overpic}[width=0.5\textwidth, trim= 0cm -1cm 0cm 0cm]{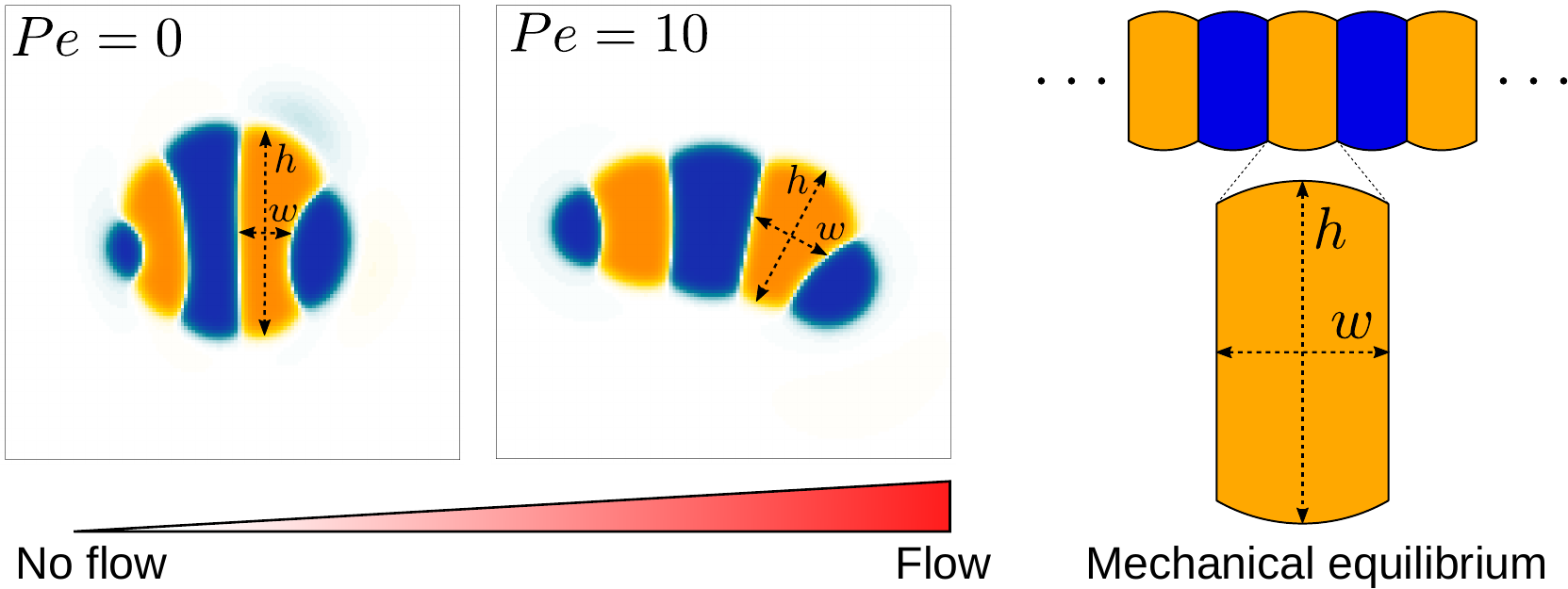}
      \put(-6,43){(g)}
      \put(63,43){(h)}
    \end{overpic}
    \input{single/AR.tex}
  \end{center}
  \caption{Transient ``croissant'' states of the droplets, which show an elongated and striped organization of the phase domains.
  (a) Illustrative example of droplet surface deformation, at $Pe=100$ and $t=0.1$, and resulting angular Fourier spectrum used to define the droplet elongation $E$.
  (b) Evolution of the droplet elongation as a function of time.
  (c) Transient droplet elongation as a function of the Peclet number at fixed numbers of phase domains, $N_D$.
  (d) Illustrative nematic field based on the $\phi$-field gradient, at $Pe=100$ and $t=0.1$.
  (e,f) Evolution of the global nematic order as a function of (e) time and (f) $N_D$, for different Peclet numbers.
  (g) Comparison for two snapshots for $Pe=0$ and $Pe=10$ with $N_D=5$.
  (h) Limiting case of infinite chain of equally-sized phase domains in mechanical equilibrium, where $h/w=2/Ca$ (see SI).
  (i) Phase domain aspect ratio $h/w$ as of function of $Pe$ for $Ca=1.5$. (i inset) Aspect ratio over $Ca$ for $Pe=10$.  The aspect ratio is averaged for time points with $N_D=5$ over all domains having exactly two neighboring domains (i.e.\ all non-branching inner domains).
  The quantities in all panels are averaged over a series of 25 simulations.
  }
  \label{fig:stripe}
\end{figure*}


We wondered whether the elongated, striped droplets that emerge for intermediate $Pe$ (Fig.~\ref{fig:snaps_Pe}) could be related to the slow down of the coarsening at intermediate $Pe$ numbers.
To test this hypothesis, we quantified the time evolution of droplet elongation $E$.  We define $E$ by expressing the droplet shape in polar coordinates and computing the second angular Fourier mode (Fig.~\ref{fig:stripe}a, details in SI).
It is normalized such that $E$ corresponds to the maximal radius variation of the second Fourier mode relative to the initial droplet radius $r_0 = L / 2$.

We find that for $Pe=0$ and very large $Pe$, the droplet elongation $E$ monotonically increases with time until $E\approx0.2$ (Fig.~\ref{fig:stripe}b), which corresponds to the elongation of the final equilibrium state.
However, for intermediate $Pe$, droplet elongation first increases as high as $E\approx0.3$ before decreasing back to the final value of $\approx 0.2$. Moreover, this regime of transient elongation is correlated in time to the phase of slow coarsening (compare Figs.~\ref{fig:speedup_Pe}d and \ref{fig:stripe}b). 
This suggests that the elongated droplet structures could indeed be related to the observed coarsening slow down for intermediate $Pe$.

Because the temporal coarsening dynamics differs when varying the $Pe$ number, we also plot the droplet elongation over the number of domains $N_D$.  This allows to compare droplet elongation $E$ at similar ``stages'' of the coarsening process across different $Pe$ numbers. We find that at the same $N_D$, elongation can be up to two orders of magnitude larger for intermediate $Pe$ (Fig.~\ref{fig:stripe}c).

We also observe in Fig.~\ref{fig:snaps_Pe} that elongated droplets show striped domain patterns.
Such a striped arrangement could indeed help to explain the observed coarsening slow down even for intermediate $Pe$.  A striped pattern of domains could be close to a mechanical equilibrium where only little hydrodynamic flows occur.  In this case, coarsening would occur almost entirely through diffusion in an Ostwald-ripening-like process: Due to an increased Laplace pressure in small stripes as compared to larger stripes of the same color, the larger domain would slowly grow at the expense of the smaller domain through diffusive fluxes across the domains of opposite color.

To test these ideas and to better understand how elongated droplet shapes emerge, we quantify the nematic order of inner domain boundaries (i.e.\ between droplet domains 1 and 2) by an order parameter $Q$ (Fig.~\ref{fig:stripe}d). 
To this end, we first introduce a tensor field $A_{\alpha\beta}$ that characterizes the local orientation of $\phi$ interfaces (Fig.~\ref{fig:stripe}d):
\begin{equation}
  A_{\alpha\beta} = (1 - \psi) \dfrac{\partial \tilde{\phi}}{\partial x_\alpha} \dfrac{\partial \tilde{\phi}}{\partial x_\beta}.
\end{equation}
Here, Greek letters are dimension indices and we use the normalized field $\tilde{\phi} = (C_1 - C_2) / ( C_1 + C_2) = \phi / (1 - \psi)$. We then introduce the tensor $Q_{\alpha\beta}$ as the spatially averaged, symmetric, traceless part of $A_{\alpha\beta}$:
\begin{equation}
  Q_{\alpha\beta} = \dfrac{1}{\langle A_{\gamma\gamma} \rangle}\Big\langle A_{\alpha\beta} - \frac{1}{2} \delta_{\alpha\beta} A_{\gamma\gamma} \Big\rangle.\label{eqn:definition_Q}
\end{equation}
Here, Einstein notation is used and the brackets $\langle\cdot\rangle$ denote spatial averaging over the system.  
We define the magnitude of the nematic order parameter as $Q = \sqrt{Q_{xx}^2 + Q_{xy}^2}$. When the domain interfaces are randomly oriented, nematic order vanishes, $Q=0$, while when all domain interfaces are straight and perfectly aligned, $Q=1$.

Generally, the nematic interface alignment $Q$ increases over time, and it does so more quickly for large $Pe$ numbers (Fig.~\ref{fig:stripe}e).  Note that the final equilibrium state has $Q\approx 1$, because it consists of a single straight internal interface.
Notably, if we plot $Q$ over $N_D$, we find no strong difference across $Pe$ numbers (Fig.~\ref{fig:stripe}f).  Indeed, Fig.~\ref{fig:snaps_Pe} also show striped domain patterns not only for intermediate $Pe$, but also for $Pe=0$.
But if the domain pattern is also striped for $Pe=0$, then why do we find elongated droplets only for intermediate $Pe$ (Fig.~\ref{fig:stripe}g)?

A possible reason for why $Pe>0$ droplets are more elongated is that hydrodynamic flows would drive the system closer to mechanical equilibrium, where in mechanical equilibrium the droplets might be elongated.
To test these ideas, we first analytically compute the mechanical equilibrium state for a striped droplet.
For simplicity, we focus on the limiting case of an infinite chain of equally sized domains (Fig.~\ref{fig:stripe}h).
In mechanical equilibrium with $\Delta\sigma=0$, interfaces between phases 1 and 2 are parallel straight lines, and surfaces to the external fluid 3 are circular arcs.
One can furthermore show that the aspect ratio of a single domain in mechanical equilibrium is $h/w=2/Ca$ (derivation in SI).
Thus, in a crude approximation, a droplet with $N_D$ domains in mechanical equilibrium would have aspect ratio $N_Dw/h=N_DCa/2$.

To test whether droplets for intermediate $Pe$ approach the mechanical equilibrium state, we quantified the domain aspect ratio $h/w$ in our simulations (details in SI).
We plot $h/w$ for $Ca=1.5$ depending on $Pe$ in Fig.~\ref{fig:stripe}i and compare to the theoretical solution of the infinite chain (dashed line).
Our results suggest that the droplets do indeed approach mechanical equilibrium as $Pe$ increases.

To further test these ideas, we compared domain aspect ratios also across different values of $Ca$ (Fig.~\ref{fig:stripe}i inset).  We found that for smaller $Ca$, the measured aspect ratio started to become smaller than the theoretical prediction.
This makes sense since for smaller $Ca$, the overall droplet aspect ratio also becomes smaller and so the simulated droplets deviate more strongly from the theoretical picture of a chain of equally sized domains.  Indeed, for $Ca\rightarrow 0$, the infinite-chain prediction would be $h/w=2/Ca\rightarrow\infty$. However, in this limit, there is no inner interface tension, $\sigma_{12}=0$, and the droplet shape is spherical. When dividing a spherical droplet into $N_D$ domains of equal width, the domain aspect ratio is $h/w\leq N_D$, which corresponds to an approximate upper bound for $h/w$.


Taken together, our results show that for intermediate $Pe$, transient droplet states form, which are striped and elongated.  They are close to mechanical equilibrium and coarsen only through diffusive fluxes, which makes them long-lived.


\begin{figure*}[!t]
  \centering
  \includegraphics[width=0.8\textwidth]{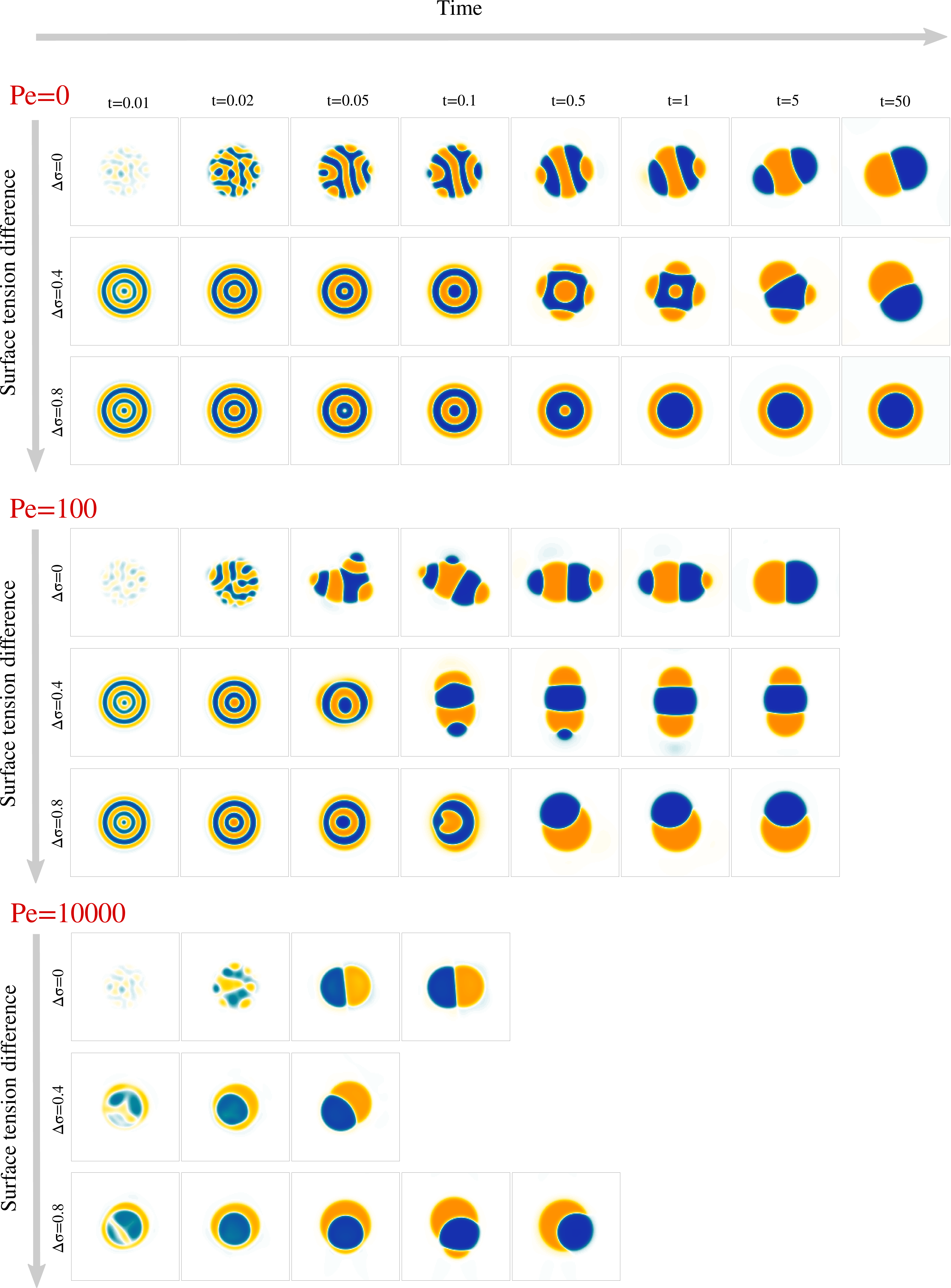}
  \caption{Surface tension difference affects transient droplet morphologies and polarization time. 
  For different values of $\Delta \sigma$ and $Pe$, a time series of snapshots is shown, representing the $\phi$ field from $-1$ (blue) to $1$ (orange).}
  \label{fig:snaps_dsig}
\end{figure*}

\begin{figure}[!t]
  \begin{center}
    \input{single/tf_dsig.tex}
    \begin{overpic}[width=0.24\textwidth]{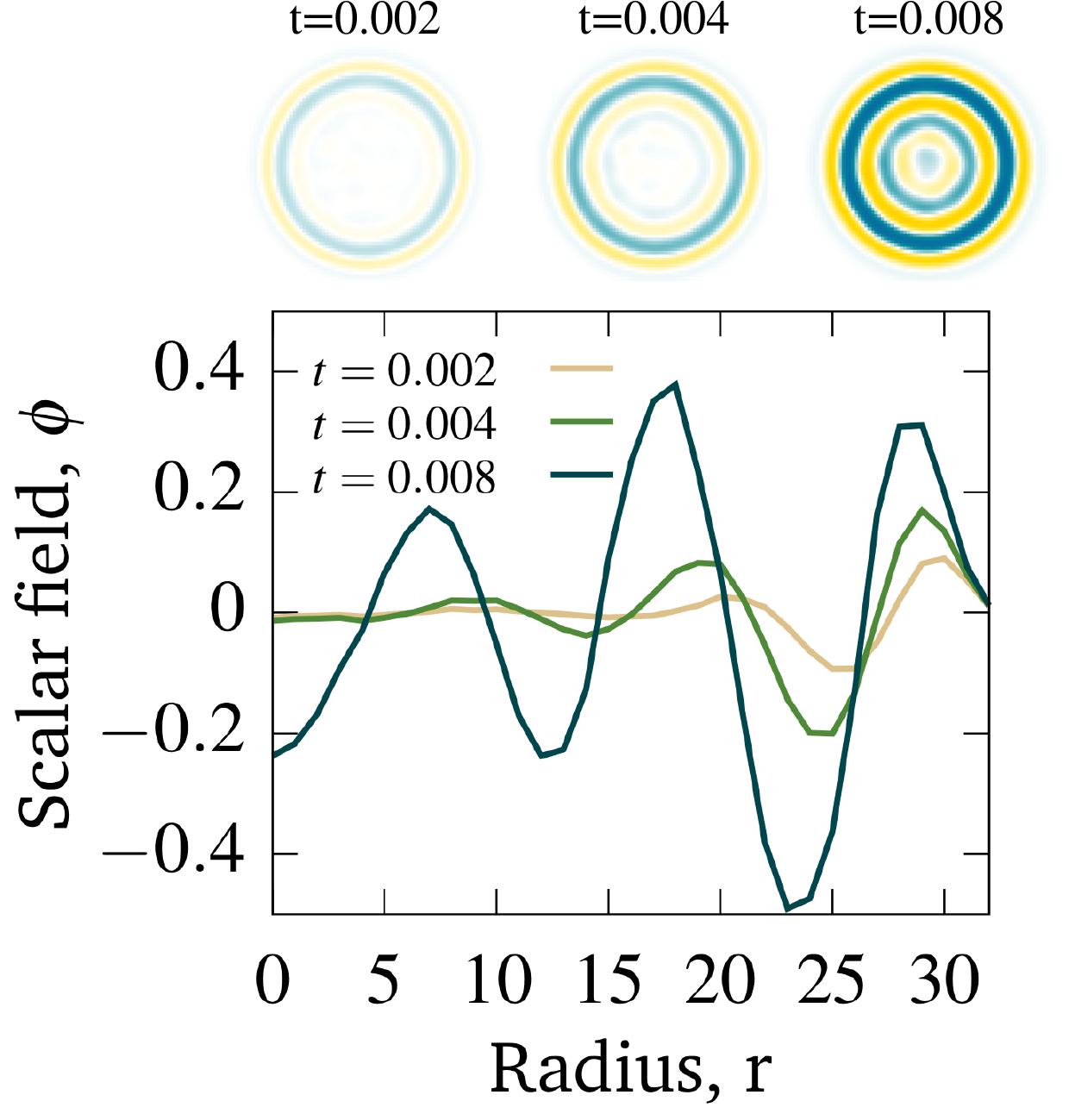}
         \put(0,80){(b)}
    \end{overpic}

    \input{single/lambda_dsig.tex}
    \input{single/nematic_tb_dsig.tex}

    \input{single/elong_te_dsig.tex}
    ~~~~~~
    \begin{overpic}[width=0.2\textwidth]{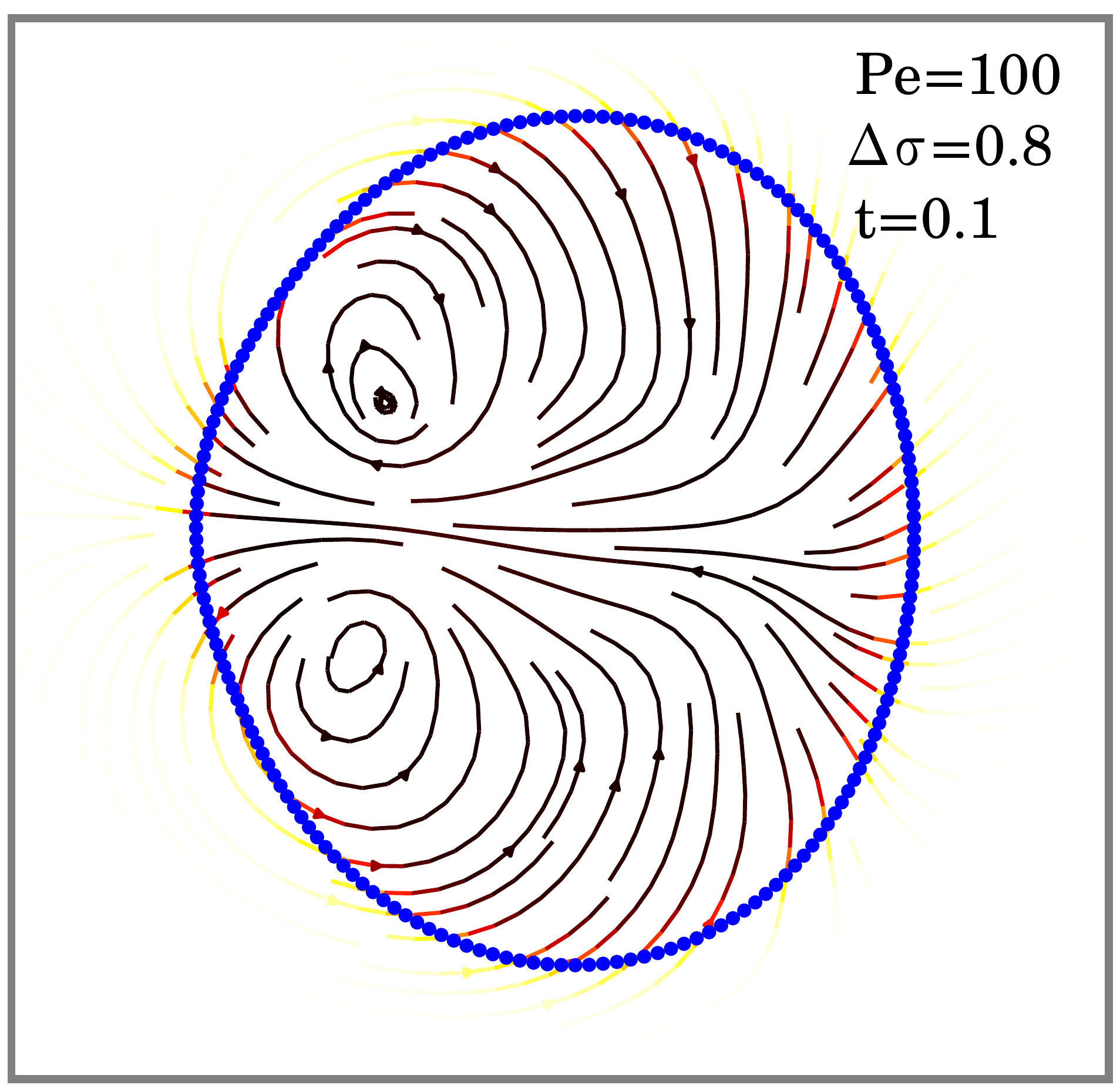}
      \put(3,85){(f)}
    \end{overpic}

  \end{center}
  \caption{(a) Polarization time as a function of $\Delta \sigma$. (b) Composition wave during the initial decomposition phase, for $Pe=100$ and $\Delta \sigma = 0.4$. (c) Domain size $\lambda$ as a function of time. (d) Symmetry breaking time $t_b$ over $\Delta\sigma$.  The symmetry breaking time $t_b$ is the first time when the nematic order $Q$ of inner domain boundaries increases above a value of 0.01. (d inset) Nematic order $Q$ of the inner domain boundaries over time for different $\Delta\sigma$. Legend same as in panel c. (e) Time $t_e$ when a croissant state (or a Janus droplet) emerges over $\Delta\sigma$. The time $t_e$ is defined as the first time when droplet elongation $E$ reaches $90\%$ of its maximum value. (e inset) Droplet elongation $E$ over time for different $\Delta\sigma$. Legend same as in panel c. (f) Directional flow during symmetry breaking.  In panels b-f: $Pe=100$.  In each of the panels c-e incl.\ insets, the respective quantity on the ordinate axis is averaged over 25 simulation runs.}
  \label{fig:dsig}
\end{figure}

\subsection{Effect of a difference between the surface tensions of the two droplet phases}
\label{sec:dsig}
We next study the effect of different surface tensions between the two droplet phases, $\Delta\sigma\neq 0$.
To this end, we carried out simulations with varying $0\leq\Delta\sigma<1$ for the Peclet numbers $0$, $100$, and $10^4$ (Fig.~\ref{fig:snaps_dsig}, Movies S1-S9).

We first examined how varying the surface tension difference $\Delta\sigma$ would change the time $t_p$ required to reach the polarized equilibrium state (Fig.~\ref{fig:dsig}a).
For $Pe=0$, only simulations with $\Delta\sigma=0$ and $\Delta\sigma=0.4$ reached the polar equilibrium state within the maximal simulation time of $t_\mathrm{max}=50$.
For $Pe=100$ and $Pe=10^4$, the equilibrium state has always been reached after a finite time $t_p$, but no clear trend in the dependency on $\Delta\sigma$ is apparent.
To better understand these results, we again separately discuss several phases of the process.

\textbf{Initial decomposition.}
For $\Delta\sigma>0$, the initial decomposition into the two droplet phases proceeds from the droplet boundary inwards in concentric circles (Fig.~\ref{fig:dsig}b).  
This phenomenon is called composition wave in the literature:\cite{lee99,tanaka01} 
For $\Delta\sigma>0$, fluid 2 (orange) has a lower surface tension, and so a stripe with an abundance of fluid 2 first appears close to the surface.  This supplants fluid 1 (blue) in this region, pushing it further inwards.  The slight abundance of fluid 1 next to the outer fluid-2 stripe attracts more of fluid 1, supplanting more of fluid 2, some of which is pushed further inwards, and so on. 
This initial decomposition proceeds until the whole droplet displays a rotationally symmetric striped pattern, where the stripes reach saturation $\vert\phi\vert\approx1$ again around $t_d\approx 0.02\tau_D$.

\textbf{Coarsening.}
The rotationally symmetric pattern coarsens through progressive broadening of the stripes and occasional annihilation of the innermost phase domain (Figs.~\ref{fig:snaps_dsig}, \ref{fig:dsig}c).
For most simulations with $Pe=0$, the system most often remained in a state with two concentric phase domains until the maximal simulation time (e.g.\ Fig.~\ref{fig:snaps_dsig}, for $Pe=0,\Delta\sigma=0.8$).
However, for finite Peclet number, this rotational symmetry got broken; for $Pe=100$ during the coarsening phase, and for $Pe=10^4$ already during the initial decomposition phase.

\textbf{Symmetry breaking.}
While there is no clear correlation of the polarization time $t_p$ with $\Delta\sigma$, we wondered whether at least the time of symmetry breaking shows a clear dependence on $\Delta\sigma$.
We thus defined the time of symmetry breaking $t_b$ as the first time when the nematic order of the inner domain boundaries, $Q$, surpassed a value of 0.01. For $Pe=100$, we found that $t_b$ increased monotonically with $\Delta\sigma$ (Fig.~\ref{fig:dsig}d \& inset).

\textbf{Transient flows.}
The symmetry breaking usually triggered complex transient hydrodynamic flows that were qualitatively similar in different parts of the parameter space, and reminiscent of Marangoni flows within a droplet (Fig.~\ref{fig:dsig}f).\cite{schmitt16,maass16}
For $Pe=100$, these flows typically resulted in a croissant state.

\textbf{Slowly relaxing ``croissant'' state.}
We wondered whether also the time $t_e$ it takes for the croissant state to emerge shows a clear dependence on $\Delta\sigma$.  We defined $t_e$ as the first time when the droplet elongation $E$ reached $90\%$ of its maximum value during any given simulation. This captures both the approximate time of maximum elongation, in cases where a croissant state was created, or the time when $E$ first reached the final plateau value, in cases where the transient flows directly created a 2-phase Janus droplet.  We find that $t_e$ does indeed increase monotonically with $\Delta\sigma$ (Fig.~\ref{fig:dsig}e \& inset).

Thus, while the time $t_p$ for the whole process varies non-monotonically with $\Delta\sigma$, the time $t_e$ when the croissant state appears increases monotonically, and is generally much smaller than $t_p$ (compare Figs.~\ref{fig:dsig}a and e).  This means that the variation in $t_p$ is due to a variation in the life time of the croissant states.  In the previous section, we concluded that for intermediate $Pe$, these states coarsen only due to an Ostwald-ripening-like process, where smaller domains shrink at the expense of larger ones. 
This suggests that the variation in $t_p$ may just be due to the way the two droplet phases happen to be distributed into different domains by the transient flows that follow the symmetry breaking.
Indeed, we find for instance that for $Pe=100$, the simulations with $\Delta\sigma=0.4$ need longest to relax to the 2-domain equilibrium state (Fig.~\ref{fig:dsig}a), and these are also the simulations where two approximately equally-sized orange domains form (Fig.~\ref{fig:snaps_dsig}), leading to an initially slow diffusion between the two domains.

Finally, in simulations with $Pe=10^4$, a first symmetry breaking event occurred at a time around $0.01$. The ensuing transient flows often resulted in a second approximately rotationally symmetric state with 2 domains at a time around $0.02$. This state only later resolved into the polar equilibrium state.
We believe this second, 2-domain rotationally symmetric state forms in this case because the first breaking of rotational symmetry occurs before the initial decomposition is finished, i.e.\ when $\vert\phi\vert$ is significantly smaller than one within the domains.  As a consequence, the interface tension between both domains, which scales as $\sim\phi^2$, is smaller, and thus the \emph{effective} value of $\Delta\sigma=(\sigma_{13}-\sigma_{23})/\sigma_{12}$ is larger than its \emph{nominal} value.
Hence, we believe that we find the rotationally symmetric 2-domain state as transient state, because the expected equilibrium state is the rotationally symmetric one as long as the \emph{effective} $\Delta\sigma$ is larger than one (Fig.~\ref{fig:intro}f).
\section{Discussion}
In this article, we discussed phase separation in a two-phase deformable droplet.
While the thermodynamic equilibrium state for such a system is well known, we studied the route to reach equilibrium.
In particular, we focused on the interplay between spinodal decomposition and advection with hydrodynamic flows created by interface tensions.
While this interplay has been studied before,\cite{Tanaka1998} it was completely unknown how it plays out in a finite, deformable system.
We found that advection can speed up the coarsening process as reported earlier.\cite{bray02} 

We moreover found that for intermediate $Pe$ numbers, elongated, striped droplet states emerge, which we call ``croissant'' states.
The stripes correspond to a nematic alignment of inner domain boundaries, which emerges for low and intermediate $Pe$. For intermediate $Pe$, the striped droplets elongate as they approach mechanical equilibrium.
These states are long-lived, because they coarsen almost exclusively via diffusion, through an Ostwald-ripening-like process.
We did not observe such ``croissant'' states for very large $Pe$ numbers, where coarsening is entirely dominated by advection, which leads to coalescence of the phase domains before a striped pattern can develop (Movie S3).

We also studied the effect of an asymmetry between the surface tensions of the two droplet phases, and we found that it changes in particular the beginning of the phase separation process. The emergence of composition waves initially creates a rotationally symmetric system, which then starts to coarsen diffusively.  Breaking of this rotationally symmetric state triggers transient flows that are reminiscent of Marangoni flows in droplets.\cite{schmitt16,maass16} These transient flows typically gave rise to a long-lived croissant state.
Taken together, our results demonstrate that advection can play an important role in the coarsening dynamics of deformable two-phase droplets.



\section{Symmetry breaking in stem cell aggregates}
This work has also important implications for the kinds of biological systems that originally motivated this study.\cite{VandenBrink2014, turner17, Beccari2018, VandenBrink2020, moris20, Hashmi2020, anlas21}
Aggregates of stem cells that mimic early vertebrate axis formation are known to develop a polar structure (Fig.~\ref{fig:intro}a,b),\cite{VandenBrink2014} reminiscent of two-phase Janus droplets in thermodynamic equilibrium (Fig.~\ref{fig:intro}d), where the different phases correspond to different cell types.
While current studies of such stem cell colonies focus almost exclusively on the biochemical dynamics,\cite{Etoc2016,Chhabra2019} our work here demonstrates that tissue flow and advection could play an important role as well. Indeed, recent experiments have identified significant large-scale tissue flows in stem cell aggregates.\cite{Hashmi2020}

As a first rough estimate, the Peclet number in mouse embryonic stem cell aggregates is on the order of $Pe\sim L\sigma/D\eta\sim200$.  This is based on an aggregate size of $L\sim200\,\micro\metre$,\cite{Hashmi2020} a tension-to-viscosity ratio of $\sigma/\eta\sim 5\,\micro\metre/\hour$,\cite{Oriola2020} and a diffusion constant of order $D\sim 5\,\micro\metre^2/\hour$ (unpublished preliminary data, lab of Pierre-François Lenne). In this $Pe$ regime, we observe in our simulations the emergence of the ``croissant'' states.  Indeed, these states appear visually reminiscent of somites, which have also been observed to form in gastruloids,\cite{VandenBrink2020} while an interface tension has been reported at somite-somite boundaries \textit{in vivo}.\cite{Shelton2021}
However, first, the mechanism of somite formation in vertebrates is generally believed to crucially involve the complex interplay of biochemical signals in a clock-and-wavefront model\cite{Oates2012} (even though purely mechanical models are discussed as well\cite{Adhyapok2021}). Second, different from our model, gastruloids exhibit a polar organization \emph{before} somites form.\cite{VandenBrink2020} This raises the question why before polarization there are usually no ``croissant'' states observed in experiments.

Stem cell cultures are of course much more complex than the model we studied here, and future refined studies should take additional effects into account.
First, experiments indicate that the dynamics of some biological signaling molecules and transcription factors is non-conserving in these systems.\cite{Etoc2016,Chhabra2019,Hashmi2020} To take this into account in our model, a source term needs to be added in the $\phi$ equation Eq.~\eqref{eqn:CH_phi}.
Similarly, we studied here a droplet of constant volume, while the stem cell colonies are typically growing. To implement this, source terms need to be added to the $\psi$ dynamics Eq.~\eqref{eqn:CH_psi} and in the incompressibility condition Eq.~\eqref{eqn:divU}.

Second, we focused here on a 2D model, while the stem cell systems motivating this study are 3D aggregates.\cite{VandenBrink2014,Hashmi2020} While there are likely several commonalities between the 2D and 3D cases, we expect differences for instance due to Laplace pressure effects.\cite{Siggia1979,tanaka01,Hutson2008} Such effects arise only in 3D, where in tube-shaped phase domains, interface tension creates a Laplace pressure that depends on the tube radius. This pressure can lead to necking and breakup of connected domains in a Plateau-Rayleigh instability,\cite{Siggia1979} but it can also lead to a pumping effect that makes the surface layer of the droplet grow more rapidly.\cite{tanaka01}  It will be interesting to dissect the consequences of these effects for a deformable 3D droplet in future work.

Third, the cell aggregates are active systems, and so activity may need to be included for an effective description of symmetry breaking in the aggregates.\cite{Tiribocchi2015a,Weber2019}

Fourth, additional hydrodynamic fields may be required to fully account for some features of the stem cell aggregate dynamics. In particular, the inclusion of more scalar fields in addition to $\phi$ may be required,\cite{Etoc2016,Chhabra2019} or of a polar field as experiments on fish aggregates have pointed to a role of polarity proteins.\cite{Williams2020,Tjhung2012a}

Finally, the real system dynamics is likely influenced by several types of noise. Moreover, a $\phi$-dependence of mechanical properties such as viscosity or local volume growth may be required to explain some of the experimental observations.\cite{Mongera2018}

\section*{Conflicts of interest}
The authors declare no conflict of interest.

\section*{Acknowledgements}
We thank Andrej Košmrlj, Pierre-François Lenne, and Sham Tlili for critical reading of the manuscript. We also thank Pierre-François Lenne and Sham Tlili for useful discussions and experimental data.
The project leading to this publication has received funding from the ``Investissements d'Avenir'' French Government program managed by the French National Research Agency (ANR-16-CONV-0001) and from ``Excellence Initiative of Aix-Marseille University - A*MIDEX''.
The Centre de Calcul Intensif d’Aix-Marseille is acknowledged for granting access to its high performance computing resources.



\balance


\bibliography{biblio} 
\bibliographystyle{rsc} 

\end{document}